# Stuck in the Turing Matrix: Inauthenticity, Deception and the Social Life of AI

Samuel Gerald Collins

Towson University, Baltimore, USA

scollins@towson.edu

## Abstract

The Turing test may or may not be a valid test of machine intelligence. But in an age of generative AI, the test describes the positions we humans occupy. Judging whether or not something is human- or machine-produced is an everyday condition for many of us, one that involves taking a spectrum of positions along what the essay describes as a "Turing Matrix" combining questions of "authenticity" with questions of "deception." Utilizing data from Reddit postings about AI in broad areas of social life, the essay examines positions taken in a Turing Matrix and describes complex negotiations taken by Reddit posters as they strive to make sense of the AI World in which they live. Even though Turing's thought experiment may not tell us much about the achievement of AGI or other benchmarks, it can tell us a great deal about the limitations of human life in the Matrix.

## Contents

**Introduction: Back to the Turing Test**

It seems clear that generative AI models are capable of passing the “Turing test” (Jones and Bergen, 2024). But as many have pointed out, the Turing test hardly seems relevant in an age of generative AI, when AI has been trained specifically to simulate human communication and produce human-quality creative content. As Pantsar (2025) writes: “Perhaps we should simply stop giving the Turing test so much attention, given its relevance in AI research seems to be increasingly limited to historical treatments.” Instead, attention has turned to efforts at resurrecting dubious measures of intelligence like the Stanford-Binet IQ test–with its long history in the eugenics movement (Gebru and Torres, 2024). Although some have proposed an expanded Turing Test that would be more capable of detecting an AI agent (Hoffman, 2022; Rahimov et al, 2025), the industry has moved on, producing a bewildering number of measures, comparisons and benchmarks for intelligence (McIntosh, 2025). But with proclamations that OpenAI had developed agents with “PhD-level” intelligence, notions of “AGI” seem as ambiguous as ever (Edwards, 2025). Perhaps the Turing Test is a fitting test for machine intelligence–and perhaps it never was (Searle, 1980). Yet there are reasons to return to it that have less to do with the intelligence of the computer than the social life it instantiates.

This essay will not enjoin the debate on artificial intelligence, artificial general intelligence, or whether or not generative AI expresses authentic human intelligence or is merely, as Bender et al (2021) have famously called it, a “stochastic parrot.” Instead, this essay explores intelligence in the context of human-AI relations: intelligence as a social interaction. As humans, we (willingly or unwillingly) take part in those negotiations every day. Our social life is imbricated at every turn with AI agents, including voting (Keller and Klinger, 2019), social media (Aiello et al, 2012) and games (De Paoli, 2017). We live in an AI world, one where we encounter countless AI systems–some of which toil in the background, recommending products, sharpening photos, while others confront us with simulacrums of human behavior and communication. Is that Facebook profile real? Is that photo real? Is that voice my son’s, or is it a phishing scam? Why was I turned down for this apartment? This job? Why was I detained by the police? In a world both supported and infiltrated by AI, every day brings a series of choices about what is real, what is simulated, what is helpful and what is harmful. We’re living the Turing test.

The Turing test is Alan Turing’s adaptation of the “imitation game”: an interrogator asks a series of questions to a person and a computer, and must decide which of two respondents is a computer. It is, as many have pointed out, a curiously indirect test, one that depends on the credulity of the human interrogator and one the capacity of the machine to deceive them (Wooldridge, 2020). Neither of these is a direct measure of intelligence–however we define it. As Pantsar (2025) remarks, “For the machine to pass the test, it needs to impersonate a human successfully enough to fool the interrogator. But this is puzzling in the wide context of intelligence ascriptions. Why would intelligence be connected as a form of deception?”

On the other hand, measuring AI through its deceptive power has the benefit of avoiding the fraught task of establishing an objective measure of intelligence altogether (Bender and Hanna, 2025; Gebru and Torres, 2024). As Wooldridge (2020: 24) points out: “Turing's brilliant move here is to sidestep all the questions and debate whether a program was “really” intelligent (or conscious or whatever).” Moreover, and despite invocations of dubiously objective benchmarks of intelligence, generative AI chatbots seem to have been developed with the Turing test in mind–as convincing simulacrums of human communication generating content that is (at least in some cases) indistinguishable from the work of humans.

But, here, it might be worthwhile to re-visit Turing’s argument. In a series of insightful articles, Sterrett identifies two tests in Turing’s seminal essay, “Computer Machinery and Intelligence.” In the first (what Sterrett calls “The Standard Turing Test”), one person sits in a room and interrogates a person and a computer both located in a separate room. “The test for machine intelligence in this second version is then how difficult it is for an ‘average’ interrogator to correctly identify which is the computer and which is the man” (Sterrett, 2000: 543). But in Turing’s essay, this “Standard Test” is preceded by an "Original Imitation Game Test” that makes an appearance at the beginning of the essay:

> “In The Original Imitation Game, there are three players, each with a different goal: A is a man, B is a woman, and C is an interrogator who may be of either gender [ . . .] C’s goal is to make the correct identification, B’s goal is to help C make the correct identification, and A’s goal is to try to fool C into making the wrong identification, i.e. to succeed in making C misidentify him as a woman” (Sterrett, 2000: 542).

At first glance, the “Standard Turing Test” would seem to involve a mere substitution. Turing certainly positions it like this: “We now ask the question, ‘What will happen when a machine takes the part of A in this game?’” (Turing, 1950).

But Sterrett makes the case that the two games are very different, and that, moreover, a Turing test drawn from the “Original Imitation Game” would make a better test of intelligence (Sterrett, 2020). Others look to this as an example of technology de-stabilizing ideologies of gender binaries (Sunden, 2015). And, of course, there have been many critiques of Turing’s thought experiment (French, 2000). But rather than focus on the question of “intelligence,” I want to look to the performative and interactive qualities of the Turing Test. This is where the differences between the two tests are most evident. The “Original” Turing Test is clearly a game of deception. Can the interrogator (“C”) be deceived into thinking that “A” is a woman? The second (“Standard”) test suggests a different dynamic. Like the “Standard” test, there is a sense of deception, but the real point of the Test is one of “authenticity”: can the Interrogator accept “A” (a computer) as a thinking agent? And, beyond this, can we, as witnesses to the test, accept the computer’s answers as signifying authentic intelligence? On the other hand, the question of authenticity isn’t really the point in the original game, where the interrogator isn't trying to recognize “A” as an authentic woman (but see Hayles, 2005). While, technically, there has only been a substitution (the computer for the man), the dramaturgy has been transformed.

The Turing test has ordinarily been taken at its face value: a thought experiment that would indicate a thinking machine. And as a thought experiment, it leaves several unanswered questions that others have considered over the last decades. How many questions should be asked by the interrogator? How long should the interaction be? Can the human interlocutor (“B”) interact with the computer and the interrogator (Rahimov et al, 2025)? But these problems with the mechanics of the Turing test are less relevant in a world of chatbots. In its social dimensions, however, the Turing test as Turing describes it would seem to have a great deal of relevance to a world suffused in generative AI models. In fact, we are faced with Turing tests all of the time. Did that assignment come from a machine or from a human? Is it ok to generate a recommendation letter for a student? Will AI replace me at my job in computer programming or data entry?

Yet these questions are asking slightly different things. If I discover that a band I follow on Spotify is the work of machines, this would seem to be an unambiguous case of deception. But if I turn around and use computer generated tracks for the “intro” and “outro” of my podcast, the question is slightly different. After all, there’s no attempt to deceive, and the big question is simply the adequacy of these 20 second filler tracks. It is doubtful if people would spend a great deal of time theorizing about the 20 second filler, while, on the other hand, there have been many reports on “AI slop” on streaming media (Knibbs, 2025). Or, perhaps, we see both of these as including different admixtures of authenticity and deception. While the Turing test may offer us little in the way of definitive evidence of machine intelligence, the dual focus on deception and authenticity is a useful frame for thinking through the dilemmas of the AI world in which we now live.

Many have suggested, as Bruce Edmunds writes, that “the Turing Test is a test of human social intelligence rather than of a putative ‘general intelligence’.” (Edmonds, 2009: 211). In other words, the Turing Test, not as a test of machine intelligence, but as a measure of the changes in the ways we think about each other because of AI–a multiagent system consisting of humans and non-humans. Even before the advent of generative AI chatbots, the prevalence of bots led to both an interest in identifying the non-human, and in ways of accommodating their behavior and interaction (Seering et al, 2018).

Much of this depends on the position we occupy in this “test.” We could be “C” (the interrogator), the person who is supposed to judge the machine-generated content. We could be “B”, the person who is a model against which to compare the computer and who, as Turing rather unhelpfully writes in his 1950 paper, should “help” the interrogator. Finally, we could be the experimenter themself: the person (or people) who have conceived and executed this elaborate experiment/game.

## The Matrix of Authenticity and Deception

Imagine a 2x2 matrix with “Inauthentic” and "Authentic" down a column, and “Deceptive” and “Non-Deceptive” along the top.

| | Deceptive | Non-Deceptive |
|---|---|---|
| Inauthentic | Inauthentic/Deceptive | Inauthentic/Non-Deceptive |
| Authentic | Authentic/Deceptive | Authentic/Non-Deceptive |

[Table 1: the Turing Matrix]

When we engage in day-to-day Turing tests, these are the inevitable questions, and the ways that we perceive content that we believe could be machine generated. There are not only 4 “basic” combinations (e.g., “Authentic/Deceptive”), but we can also graph examples to represent more nuanced understandings, a spectrum of authenticity/inauthenticity/deception/non-deception against which we measure the world and chart our course through it. Along the way, there are questions here of intentionality, of social relations, of power and hierarchy, all of which can be at least partially captured in Turing’s matrix.

## The Question of Deception

Mitchell (2024) points out that many researchers see the Turing test’s indirect measure of intelligence as its biggest liability: “Because its focus is on fooling humans rather than on more directly testing intelligence, many AI researchers have long dismissed the Turing Test as a distraction.” Yet there is little doubt that many of the AI models that people encounter in their everyday life is in the context of deception: deep fakes, fake books, machine-generated music, fake pictures, social media profiles, fake students, fake employees and so on (Schmitt and Flechais, 2025). All of this has driven an AI-detector industry–with widely disparate success rates (Malik and Amjad, 2025). There is more to deception, though, than just a willingness to deceive (Natale, 2023). Natale identifies a spectrum of deceit, from “explicit attempts to commit fraud, mislead and tell lies” to what he calls “banal deception,” which "includes the use of humanized, gendered voices of AI assistants such as Amazon’s Alexa and Apple’s Siri” (Natale, 2023). In the case of “banal” deception, no fraud is intended; we know that Alexa is not a real (gendered) person, but still accept the deceit as an effective(?) interface. Ultimately, Natale makes the argument that deceit has been inseparable from the growth of AI (Natale, 2021). And yet, the AI World is not always just (or primarily) about deceit.

**The Question of Authenticity**

The Turing test is, in the end, supposed to indicate authentic intelligence–or rather than its simulacrum. Whether or not it is capable of this has been a lengthy debate in the decades since Turing’s essay, with plenty of skeptics (e.g., Searle’s Chinese Room) critiquing the test as a valid measure (Searle, 1980). But, as a social practice, our measure of AI depends upon some criteria of what constitutes “authentic” human work and whether or not AI models are able to generate this level of authenticity. And therein lies the tension, since most of us would, in this context, identify “authentic” with “human,” and machine generated with “artificial” and “inauthentic.” As Ressler (2025: 985) writes, “the technology embodies inauthenticity in a way other technologies do not.” In practice, though, people encounter AI-generated content without that level of philosophical critique. Yet while users may not be primarily concerned with AI as a genuine intelligence, they nevertheless consume and utilize the products of AI in the context of authenticity and inauthenticity. Is the code generated by Github Copilot as good as that written by humans (Imai, 2022)? Is the internet doomed to a spiral downwards into “AI Slop,” or is it possible that AI-produced media will one day be acceptable to humans as possessing “authentic” qualities (Bellafiore, 2025)? What about the writing and media content we generate to save time and (purportedly) to increase productivity? Can we accept these as “authentic” in the workplace? Questions of authenticity extend to other AI systems as well, including recommender systems, algorithmic scoring of job applications, facial recognition systems. Are these producing “authentic” results or not? As in the case of deception, questions of authenticity very much depend on our position in the dramaturgy of the Turing test. Are you taking the position of “interrogator” in the case of someone in your organization trying to introduce machine-produced work as "authentic"? Or, are you taking the position of the computer and adopting AI models in order to accomplish more work? Or are you in the position of “B”--producing work that is then compared to machine-produced content, and wondering perhaps about the future of your job?

**Methods**

The Turing test, and the Matrix of Authenticity and Deception it implies, then, reflects social conditions in our AI world. In order to explore and illustrate this, I have utilized NodeXL to download three sets of Reddit data through Reddit's API that discuss AI in very different milieux. There are, of course, many subreddits that concern AI, including popular ones like "ChatGPT," "GPT3," "ArtificialIntelligence" and many other variations on these. But to explore the meanings of AI in social life, it was important to examine subreddits that were not outwardly concerned with AI, and the three subreddits considered here range widely in topic through multiple themes in work, education and dating. Next, the posts were then analyzed according to the frequency of word pairs, the results graphed by related groups using social network analysis (Smith et al, 2010). The resulting graphs show both thematic clusters and dense linkages between groups. Although many of the people posting to subreddits considered here are critical of AI, the analysis seeks to highlight the ways in which people negotiate the Turing Matrix and occupy various positions within it as they strive to make sense of an AI world and accommodate themselves to it.

**Analysis of Subreddits**

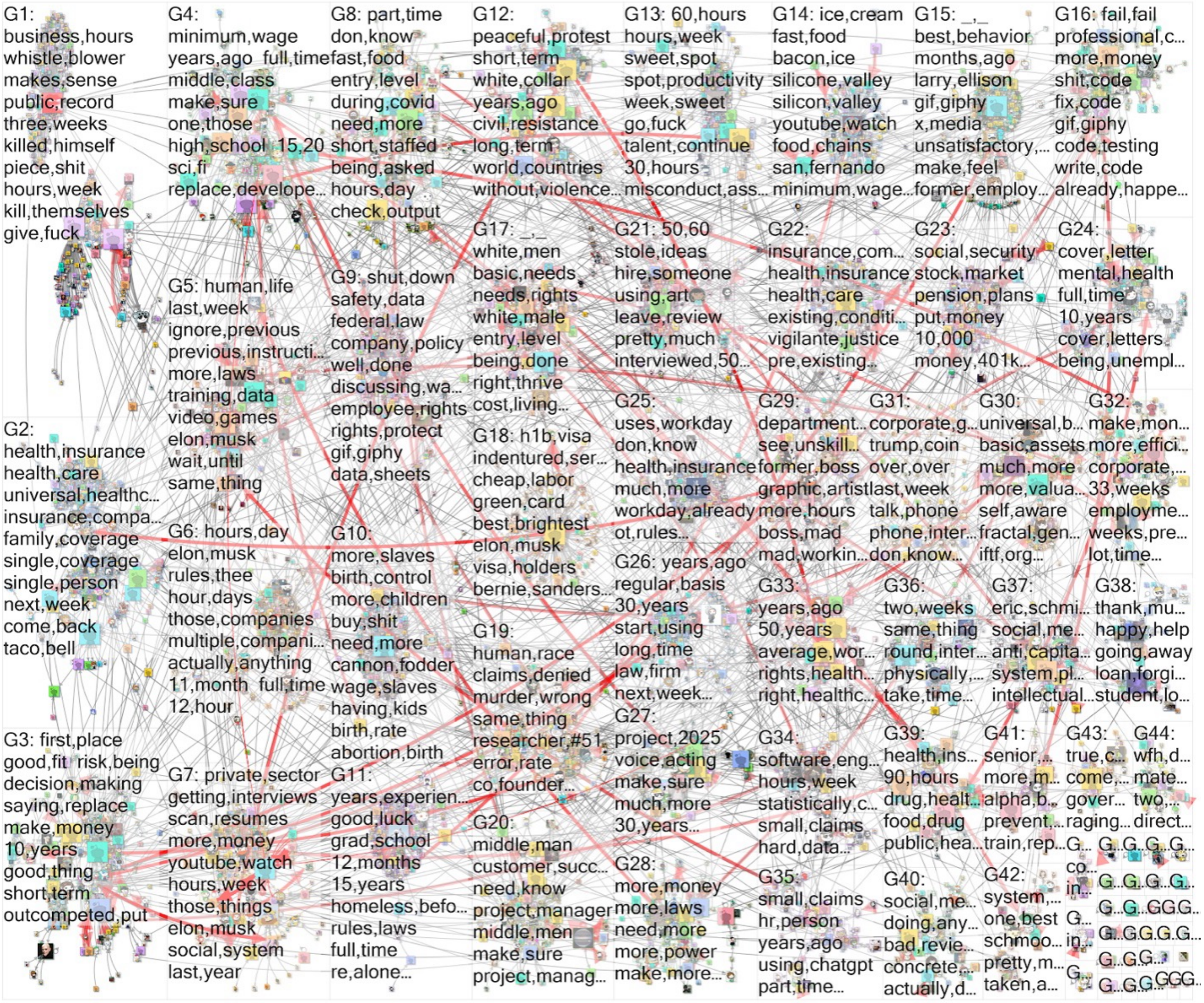

[Figure 1: NodeXL graph showing posts on AI from the “antiwork” subreddit]

The “antiwork” graph includes 8263 posts, comments and replies on AI from June of 2024 to March of 2025. Clustered in groups using the Clauset-Newman-Moore algorithm, the graph shows the 10 most common word-pairs for each group. As a subreddit that critiques work, it is not surprising to find that discussions of AI center on its negative impacts on the workplace. Top posts include fears over job replacement by AI, complaints about the added work AI-produced code and content gives employees, and critiques of CEOs and HR departments that utilize AI in hiring, firing and downsizing the workforce. More positive postings concern the potential of AI tools to allow workers more leisure time.

I have thematized posts and comments for the 10 largest groups:

G1: The suspicious death of whistleblowers at Boeing.

G2: Replacement of humans with AI in computer coding/ the failure of AI to replace humans in coding.

G3: Getting replaced by AI at work, and having to correct the mistakes introduced by AI.

G4: AI being used to fire workers because of the greed of billionaires.

G5: AI making hiring decisions.

G6: AI leading to overwork.

G7: AI reading resumes; using AI to create resumes, cover letters.

G8: Employers not hiring workers, despite posting jobs online.

G9: AI increases in profit aren't shared with workers.

G10: AI detectors in the workplace.

| | Deceptive | Non-Deceptive |
|---|---|---|
| Inauthentic | | G5<br>G3,G6<br><br>G2 |
| | | |
| Authentic | G10 G7<br><br>G4<br>G9 | |

[Table 2: NodeXL groups distributed on the deception/authenticity 2x2 matrix]

The matrix plots each of the groups according to their position vis-à-vis deception and inauthenticity. While one could certainly debate whether something is more or less deceptive, or more or less authentic, it’s clear that not all examples of “authentic” describe the same levels of authenticity. For example, using an AI detector straddles the line between “authentic” and “inauthentic,” “deceptive” and “non-deceptive.” While deployed in a relatively transparent way by workplace management, detectors police all of those boundaries, separating authenticity from inauthenticity, deceptive from non-deceptive. On the other hand, being forced to utilize AI at work for programming is non-deceptive, but ultimately inauthentic—since Reddit posters claim that their AI-generated code doesn’t work, and that they have to go through laborious debugging. Group 1, although featuring some commentary on AI, dealt almost exclusively with the fate of Boeing whistleblowers.

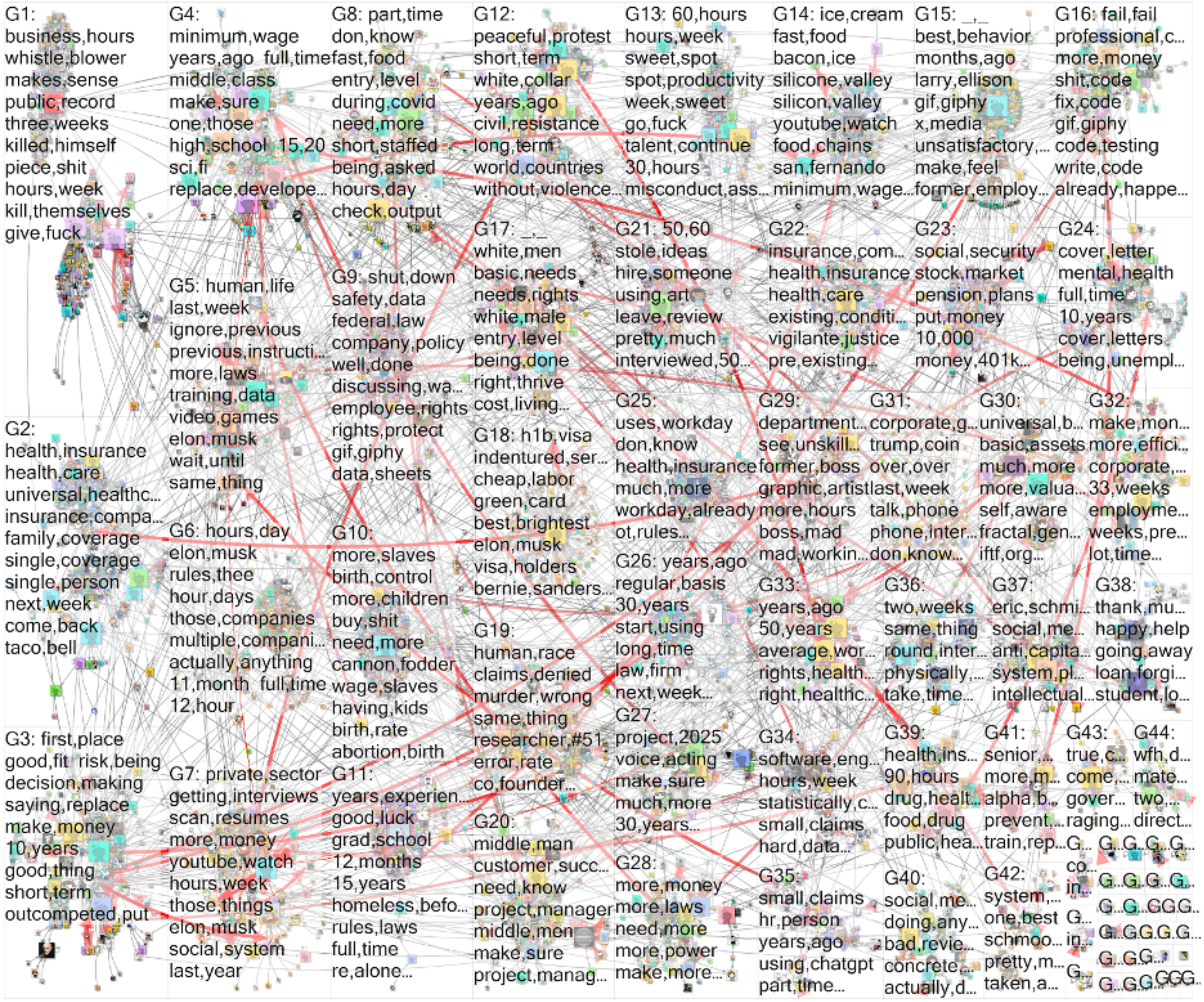

[Figure 2: NodeXl graph showing posts on AI from the “professors” subreddit]

The “professors” graph collected 4426 posts and comments on the subject of AI appearing in the “professors” subreddit from December of 2024 to March of 2025.

Clustered in groups using the Clauset-Newman-Moore algorithm, the graph shows the most common word pairs for each group. Since most of the commentary on the “professors” subreddit is critical about higher education in general, it is not surprising that most redditors complained about the impact generative AI tools were having on their classes, many expressing considerable dismay over the high percentages of AI-produced assignments in their courses. They strategized ways to minimize AI usage in their classes, including adopting syllabus language, changing assignments to more in-class work, and "catching" students at their AI usage. The comparatively small number of more positive comments looked at ways AI tools might make assignments more interesting, or assist in the evaluation of student work. A thematic analysis of posts and comments for the 10 largest groups suggests these overlapping concerns, with the exception of “Group 1,” which focused on a video link.

G1: Commentaries on a video link.

G2: Difficulties of teaching writing classes when all of the students are using AI.

G3: Crafting syllabus language, rubrics and assignments to prevent AI misuse.

G4: AI generated emails from students and ways to improve student communication.

G5: Students can’t learn with AI, and leave school without knowing how to think.

G6: Giving oral exams and the problem of loading student data to AI detectors.

G7: New ways to cheat using AI - with examples from classes (with student names removed).

G8: The ways technology has diminished the abilities of students.

G9: Students unable to see the value of critical studies of race, gender and class.

G10: Teaching and assessing composition courses in light of AI.

| | Deceptive | Non-Deceptive |
|---|---|---|
| Inauthentic | G7<br><br>G4 | G5<br>G6<br>G8, G9<br><br>G10 |
| Authentic | G2 | G3 |

[Table 3: Top groups of Professors on a 2x2 matrix]

Cheating on course assignments (G7) seems both “inauthentic” and “deceptive,” but writing assignments to discourage students from cheating (G2) acknowledges the deception while attempting to coax authentic work from a class. The many commentaries on the cognitive and communicative impacts of AI (G5, G8, G9) are meditations on inauthenticity in contemporary life, but are not concerned with cheating per se. On the other hand, and perhaps because this subreddit is for college faculty, the crafting of syllabus language to discourage AI use (G3) appears here as both “authentic” and “non-deceptive.”

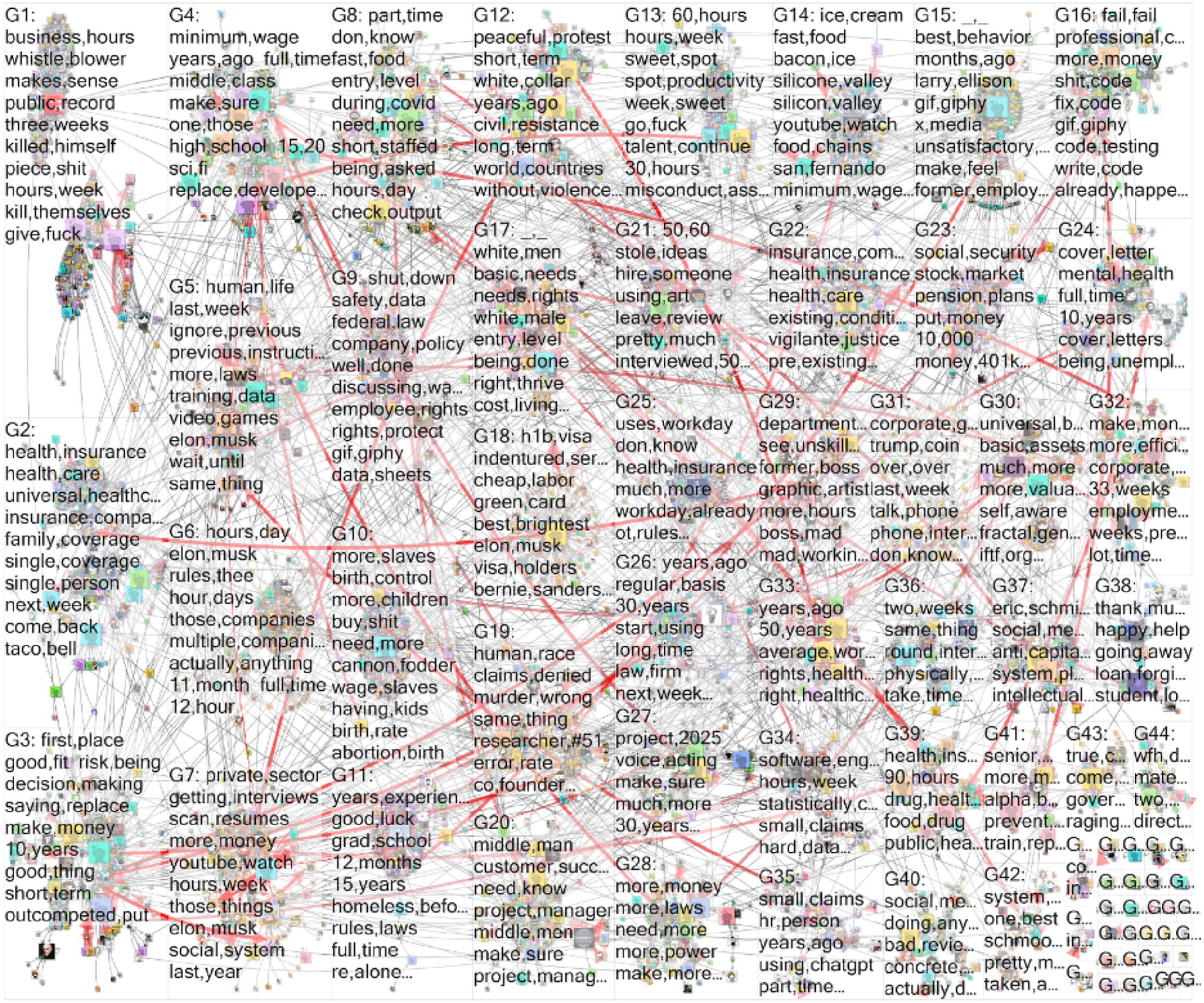

[Figure 3: Posts on AI from the “dating” subreddit]

The “dating” graph collected 2716 posts and comments on the subject of AI appearing in the “dating” subreddit from October of 2017 to March of 2025. Clustered in groups using the Clauset-Newman-Moore algorithm, the graph shows the most common word pairs for each group. Of the three subreddits, this one featured the smallest number of posts, and, perhaps because of this, the API returned results going back to 2017. “AI” (however construed) has been an issue in online platforms for some time, with AI-generated profiles, allegations of bias in algorithms, and automated banning being a source of concern for these redditors before 2022. Since the advent of ChatGPT in 2022, other issues have surfaced, including the ethics of using chatbots to practice texting or generate responses in text-based conversations.

G1: Men using AI to generate texts to talk to women.

G2: Using AI to produce and/or edit your profile.

G3: Getting banned by the app’s algorithm.

G4: Difficulty in dating in real life.

G5: Being unemployed and keeping your job profile secret.

G6: Dating people who are politically opposite you.

G7: Using an AI-powered “delusion calculator” to show the impossibility of finding a “perfect” match.

G8: Digital versus real life interaction, with respect to bots.

G9: Practicing digital interaction with an AI chatbot.

G10: Having nerdy hobbies and being passionate about intellectual things.

| | Deceptive | Non-Deceptive |
|---|---|---|
| Inauthentic | G8, G1, G2<br><br>G5 | G9 |
| Authentic | G7 | |

[Table 4: Top groups from the “dating” subreddit]

Not all of the groups dealt primarily with AI, automation or algorithms, but AI nevertheless came up in passing in posts in other groups. Only those groups that dealt primarily with AI are represented in the matrix above. Debates over authenticity and deception tended to be framed as ethical quandaries. Should you use AI to text potential dates? And what happens when you meet in real life and you're different from your AI doppelganger? On the other hand, there were also posts about trying to interact with bot profiles that weren't linked to a real person at all. Finally, many posts concerned the use of chatbots to "practice" relationships.

## Discussion

This brief survey of Reddit posts tells us something about human-AI interaction today. Rather than merely an acceptance or rejection of AI in social situations, people take a variety of positions as they grapple with the AI in their lives. "Authenticity" and "deception" are negotiated, and that negotiation takes very different forms, forms that are less about a consensus about what "intelligence" might be with AI, and more about moments when people feel most conflicted. Are people on the receiving end of AI content? Are they making it themselves? Do they feel it's making their lives easier or more difficult? Are they using AI to benefit themselves, or the organization in which they work? Is it deceiving them? Or is it helping them to live better lives? Although beyond the scope of this brief survey, it seems reasonable that people would take different positions on AI depending on the context. For example, the same chatbot that might be helping you practice your interactions with potential dates might also be extruding buggy code that is making your life more onerous even as it's introduced as a boon for productivity.

In this respect, generative AI seems very similar to interactions with bots that have been ubiquitous in social media and in RPGs for a number of years—i.e., not really about "acceptance" or "rejection" than about negotiation and accommodation. "The question is no longer whether bots can pass, but how social interaction with them may be meaningful" (Jones, 2015: 1). And like bots, "AI" is hardly monolithic, taking multiple forms even in the same application. But AI has become ubiquitous in many parts of our lives. So it is hardly surprising that reactions are ambivalent.

The other part of this is the location of the humans in these "tests": the position of "A," "B" or "C" in these Turing Tests. Are you in a position where you are wondering whether something is real or fake? That is, in the position of the interrogator? Are you in a position where you are trying to pass something off as your work when it's been extruded by machine? That is, taking the position of the computer? Are you faced with conditions where someone in your organization is trying to introduce machine-produced work as "authentic"? Or someone who has been left to deal with a bunch of AI Slop (content, media content, etc.), and it's your job to make sense of it all. That is, are you in the position of the other human (B)?

**Conclusion**

What does it matter whether or not we're stuck in the Turing Matrix? Simply this: in the matrix, other alternatives outside of the matrix are closed to us. The Turing test was a thought experiment, but the Turing Matrix describes the ontological and epistemological challenges we face every day. Every phone call, every text, every online encounter, every news story, photograph, film, piece of music: our day is in the Turing Matrix, parsing between authenticity and deception, bad actors, unreliable information, machine content, human content.

To be stuck in the matrix is to be stuck in the position of one of the humans in this matrix - and to be less human because of it. As Beerends and Aydin (2025) conclude, “humans are reduced to information processing stimulus-response machines and the brain is considered to be similar to a digital computer.” From this perspective, the important part in the Turing Matrix is not really the putative intelligence of the computer, nor is it the credulity of the interrogator. Instead, the Turing Matrix circumscribes the human in order to elevate the artificial. To succeed, people must be convinced that algorithmic processes are interchangeable with the etiolated human. In other words, the Turing test only works if we stay in our room, if we don’t shout, if we don’t bang on the walls with our fists. Instead, we’ll need to reduce ourselves to the computer in order for this “test” to work. And there are 2 levels of deception in this maneuver- one in convincing (or forcing) people to reduce themselves to narrowly defined outputs, the other in misrecognizing those algorithmic products as the total of human possibility.

Of course, people don’t readily accede to these circumscribed roles. This brief analysis of Reddit postings shows people chafing against the Turing Matrix in various ways: questioning, unmasking the powerful interests lurking behind a putatively neutral technology, and critically reflecting on the ultimate impacts of all of this on their work and lives. Yet, whether critical or salutary, reactions still slot into the Turing Matrix. This is one of the inescapable conditions of living in an AI World—being forced to evaluate deception and authenticity, yes, but only from the terms of the machine. That is to say, assessments of deception and authenticity are—by default—disembodied (Hayles, 1999). We make these determinations as Turing intended, from clues in the content itself rather than by, say, observing a person write, draw or perform music.

There are alternatives to all of this, from Zuboff's call to collective action as a challenge to an extractivist, surveillance capitalism (Zuboff, 2019), to Ben Schneiderman's more mild prescriptions for a human-centered AI (HCAI) where we remain in control of technologies that we use for the good of our communities (Scheiderman, 2022). And we can pursue alternatives at any time. We can question the authenticity of machine-generated content. We can give up the powerful, harmful fictions that fuel the Matrix: that machines can be built that display human intelligence, and that this would be a desirable goal. After all, As Bender and Hanna (2025: 189) conclude, "At no point, however, does calling any of this technology "AI" help. The term obscures how systems work (and who is using them to do what) while valorizing the position and ideas of power brokers." Finally, we can take the Turing Matrix to heart, and realize that our AI World is, literally, the production of a world of human and non-human agents, one in which we–the humans–can control the terms of the Test. It only works if we comply with the rules, and the moment we begin to question them, the Matrix gives way.

**About the Author**

Samuel Gerald Collins is a professor of anthropology at Towson University. His research includes urban studies, social media, design anthropology, and information technologies in South Korea and in the United States. He is the author of *All Tomorrow's Cultures: Anthropological Engagements With the Future*, and co-author (with Matthew Durington) of Multimodal Methods in Anthropology.

Acknowledgments

**Tables and Figures**

[Table 1: the Turing Matrix]

-See attachment

[Figure 1: NodeXL graph showing posts on AI from the “antiwork” subreddit]

| | Deceptive | Non-Deceptive |
|---|---|---|
| Inauthentic | | G5<br>G3,G6<br>G2 |
| Authentic | G10 G7<br>G4<br>G9 | |

[Table 2: NodeXL groups distributed on the deception/authenticity 2x2 matrix]

-See attachment

[Figure 2: NodeXl graph showing posts on AI from the “professors” subreddit]

| | Deceptive | Non-Deceptive |
|---|---|---|
| Inauthentic | G7<br>G4 | G5<br>G6<br>G8, G9<br>G10 |
| Authentic | G2 | G3 |

[Table 3: Top groups of Professors on a 2x2 matrix]

-See attachment

[Figure 3: Posts on AI from the “dating” subreddit]

| | Deceptive | Non-Deceptive |
|---|---|---|
| Inauthentic | G8, G1, G2<br><br>G5 | G9 |
| Authentic | G7 | |

[Table 4: Top groups from the “dating” subreddit]